\documentclass[preprint,
 amsmath,amssymb,
 aps,
]{revtex4-1}

\usepackage{graphicx}
\usepackage{dcolumn}
\usepackage{bm}
\usepackage{multirow}

\begin{document}

\title{Guided modes and terahertz transitions for two-dimensional Dirac fermions in a smooth double-well potential}


\author{R. R. Hartmann}
\email{richard.hartmann@dlsu.edu.ph}
\affiliation{
Physics Department, De La Salle University,  2401 Taft Avenue, 0922 Manila, Philippines
}

\author{M. E. Portnoi}
\email{M.E.Portnoi@exeter.ac.uk}
\affiliation{Physics and Astronomy, University of Exeter, Stocker Road, Exeter EX4 4QL, United Kingdom}
\affiliation{ITMO University, St. Petersburg 197101, Russia}

\begin{abstract}
The double-well problem for the two-dimensional Dirac equation is solved for a family of quasi-one-dimensional potentials in terms of confluent Heun functions. We demonstrate that for a double well separated by a barrier, both the energy level splitting associated with the wavefunction overlap of well states, and the gap size of the avoided crossings associated with well and barrier state repulsion, can be controlled via the parameters of the potential. The transitions between the two states comprising
a doublet, as well as transitions across the pseudo-gaps are strongly allowed, highly anisotropic, and for realistic graphene devices can be tuned to fall within the highly desirable terahertz frequency range.
\end{abstract}

\maketitle

\section{Introduction}
The double-well in quantum mechanics has been studied in relation to various physical phenomena, ranging from vibrations of polyatomic molecules~\cite{rosen1932vibrations}, through to applications in Bose–Einstein condensation~\cite{milburn1997quantum}. Solutions to the Schr{\"o}dinger equation for smooth double-wells are equally broad, and have been analyzed using perturbative methods~\cite{muller2012introduction}, instanton calculus \cite{gildener1977pseudoparticle},   the WKB approximation~\cite{landau2013quantum} and other techniques.
With the rise of two-dimensional (2D) Dirac materials~\cite{wehling2014dirac}, comes a fresh opportunity to revisit the  double-well problem in a relativistic setting, and to conduct ultra-relativistic 
experiments without the need for powerful accelerators. Indeed, there has been significant progress in creating guiding potentials in Dirac materials~\cite{huard2007transport, ozyilmaz2007electronic, gorbachev2008conductance,liu2008fabrication,williams2011gate,rickhaus2015guiding}, most recently
using carbon nanotubes as top gates~\cite{cheng2019guiding}. These top-gated structures allow the potential profile to be controlled by manipulating the top-gate voltage, allowing the creation of well-defined smooth double wells.

Several approaches have been considered to achieve the goal of confinement in Dirac materials. These range from utilizing magnetic fields~\cite{de2007magnetic,masir2009quasibound,roy2012localization,downing2016massless,downing2016magnetic}, to implementing Fermi velocity engineering~\cite{peres2009scattering,raoux2010velocity, concha2010effect,downing2017localization}, through to introducing a spatially-dependant mass term~\cite{downing2019trapping}, and most commonly, using electrostatic potentials~\cite{pereira2006confined,cheianov2007focusing,silvestrov2007quantum,tudorovskiy2007spatially,shytov2008klein,beenakker2009quantum,zhang2009guided,matulis2008quasibound,bardarson2009electrostatic,zhao2010proposal,he2014guided,recher2010quantum,hartmann2010smooth,downing2011zero,stone2012searching,hartmann2014quasi,downing2014one,hasegawa2014bound,xu2015guided,xu2016guided,downing2015optimal,lee2016imaging,hartmann2017two,bai2018generating}. Confinement in massless Dirac materials is notoriously difficult due to the Klein tunneling effect~\cite{klein1929reflection,katsnelson2006chiral}. However, total confinement can be achieved in such systems for zero-energy states within an electrostatically defined waveguide, whose potential vanishes at infinity. This is because the density of states vanishes outside of the waveguide~\cite{hartmann2010smooth, hartmann2014quasi, downing2014one, hartmann2017two}. For non-zero-energy states, the bound states within the potential can couple to the continuum states outside the channel, and are therefore poorly guided. However, for massive particles, bound states can occur in spite of the Klein phenomena. 

\begin{figure}[ht]
    \centering
    \includegraphics[width=0.4\linewidth]{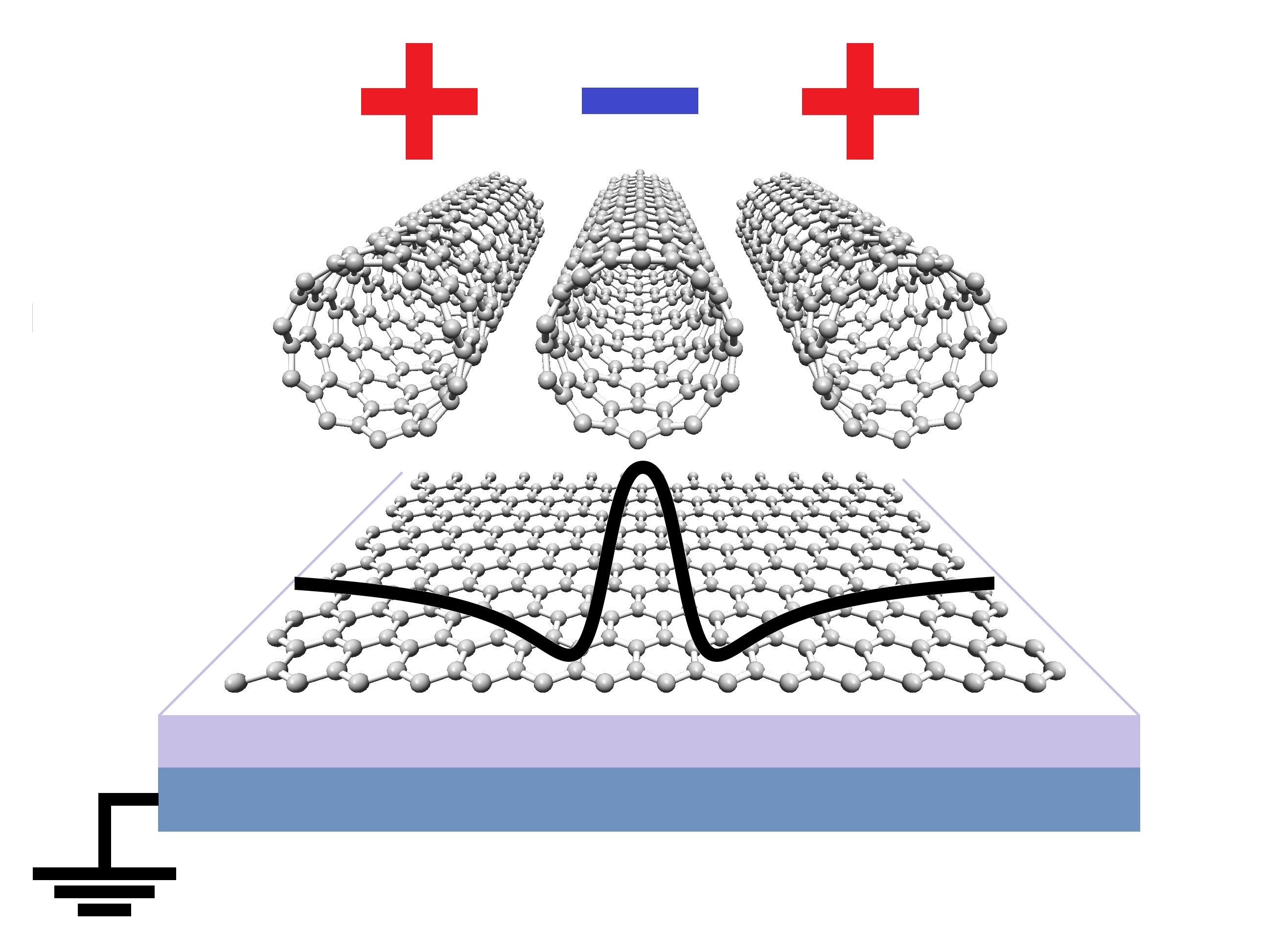}
    \caption{A schematic diagram of a double-well system, created by three carbon nanotube top gates. The central nanotube is negatively biased to create the central barrier, while the other two are positively biased to create the two wells. The Dirac material sits on top of a dielectric layer (violet layer), which lays on top of the metallic back gate. The Fermi level can be controlled using the back gate potential. The electrostatic potential created by the top gates is shown by the thick black line. }
    \label{fig:my_label}
\end{figure}

The alternative geometry of transmission through potential barriers has also been a subject of extensive research, with the majority of studies utilizing ``sharp but smooth potentials'', i.e., potentials which are step-like or have kinks but are assumed to be smooth on the scale of the lattice constant, so that the effects of inter-valley scattering are neglected. Supercritical transmission~\cite{adler1971relativistic,dong1998relativistic,dombey1999seventy,dombey2000supercriticality,kennedy2002woods,villalba2003transmission,kennedy2004phase,guo2009transmission,miserev2016analytical} and tunneling through barriers has been studied for a variety of one-dimensional (1D) model potentials in both massless and massive 2D Dirac systems 
~\cite{roy1993relativistic, katsnelson2006chiral,cheianov2006selective,jiang2006low,bai2007klein, tudorovskiy2007spatially,cheianov2007focusing,barbier2008dirac, nguyen2009resonant, barbier2009dirac, villalba2010tunneling, pereira2010klein, xu2010resonant, miserev2012quantum, moldovan2012resonant, reijnders2013semiclassical, fallahazad2015gate, hsieh2018electrical, avishai2020klein, brun2020optimizing} including square barrier structures such as the double barrier~\cite{pereira2007graphene, pereira2008resonant, wei2013resonant}, inverted double well~\cite{bahlouli2012tunneling,alhaidari2012relativistic}, 
asymmetric waveguides~\cite{he2014guided, xu2016guided} and various other step-like structures~\cite{alhaidari2012relativistic, hasegawa2014bound, xu2015guided}. A variety of approaches ranging from the transfer matrix method through to the WKB method have been used. However, there is a dearth of studies concerning potentials which span both positive and negative energies, i.e., contain both electron-like and hole-like guided modes. The exceptions are periodic potentials~\cite{brey2009emerging} and sinusoidal multiple-quantum-well systems~\cite{xu2010resonant}. 
Multiple-quantum-well systems are shown to exhibit transmission gaps in the electrons and holes spectra at tilted angles of incidence~\cite{xu2010resonant,pham2015tunneling}, and the number of oscillations in the transmission window depends on the number of quantum wells. For a double well studied in this paper (see Fig.~\ref{fig:my_label} for the Gedankenexperiment sketch), the width of the transmission gap simply corresponds to the energy difference between guided modes, whereas the oscillations are associated with the splitting of a guided mode into multiple modes due to tunneling between wells. In this work, we consider the regime where the potential results in an ``inversion'' of electron and hole states, i.e., the valence band states in the barrier are higher than the conduction band states in the two wells.  In this regime, the 
repulsion of electron and hole states gives rise to interesting features in the eigenvalue spectrum, namely avoided crossings, which can be controlled via the applied potential. 
These level avoided crossings, 
also provide a clear physical picture behind the anisotropic energy gap opening near the Dirac point of graphene subjected to a periodic potential~\cite{brey2009emerging,wang2014transient}. 




The 2D Dirac equation, 
has been the subject of considerable interest due to the explosion of research in transition metal dichalcogenides (TMDs)~\cite{xiao2012coupled}, Weyl semimetals~\cite{young2012dirac}, topological insulators~\cite{hasan2010colloquium, qi2011topological,zazunov2016low}, low-dimensional forms of carbon~\cite{neto2009electronic,hartmann2014terahertz} and their silicon analogs~\cite{liu2011low}, for which their low-energy excitations can be described by a Hamiltonian of the form 
\begin{equation}
\hat{H}=\hbar v\left(\hat{k}_{x}\sigma_{x}+s_{\mathrm{K}}\hat{k}_{y}\sigma_{y}+k_{z}\sigma_{z}\right),
\label{Dirac_eq}
\end{equation}
where $\hat{k}_{x}=-i\frac{\partial}{\partial x}$, $\hat{k}_{y}=-i\frac{\partial}{\partial y}$,
$\sigma_{x,\,y,\,z}$ are the Pauli spin matrices, $v$ is the Fermi velocity which plays the role of the speed of light, $s_{\mathrm{K}} = \pm 1$ is the valley index number, and $k_z$ is proportional to the particle's in-plane effective mass. For graphene, $v = v_{\mathrm{F}} \approx 10^{6}$  m/s, and $k_z = 0$~\cite{wallace1947band}. For quasi-1D forms such as narrow-gap carbon nanotubes and certain types of graphene nanoribbons~\cite{dutta2010novel,chung2016electronic}, the operator $\hat{k}_y$ can be substituted by the number $k_y = E_g /( 2\hbar v_{\mathrm{F}})$, where $E_g$ is the value of the bandgap, which can be manipulated via the application of external fields~\cite{portnoi2008terahertz, hartmann2010optoelectronic, hartmann2011excitons, hartmann2015terahertz, hartmann2014terahertz, hartmann2019interband}. Examples of massive 2D Dirac systems include, but not limited to silicene, germanene, TMDs, and graphene on top of lattice-matched boron nitride~\cite{giovannetti2007substrate}.

The gapless spectrum of graphene, and the nearly gapless band structure of narrow-gap carbon nanotubes and ribbons caused a natural attraction to their optical properties in the terahertz (THz) spectral range, which have led to a menagerie of promising applications in the field of THz optoelectronics~\cite{hartmann2014terahertz}. The main goal of this paper is to demonstrate that the gate-induced double-well geometry allows for tuneable THz transitions between the various guided modes. 
We approach this problem by calculating the dispersion of guided modes in several model potentials which are smooth at the atomic scale thus allowing us to disregard the problem of valley mixing caused by jumps and kinks in piecewise potentials. These models describe well the shape of an electrostatic potential created in plane by three differently charged nanowires placed above a metallic gate, as can be shown by the mirror charge method.  

The detailed calculations of the energy spectrum for a smooth double well, with the middle barrier exceeding the side barriers are provided in Section~\ref{spectrum}, whereas the model potential with a middle barrier below the side barriers is treated in Appendix~\ref{tanh}. The selection rules for dipole transitions between the guided modes in the potential described in Section~\ref{spectrum} are analyzed in Section~\ref{Transitions} followed by the summary of the results in Section~\ref{Conclusions}.


\section{Relativistic one-dimensional double-well problem}\label{spectrum}
In what follows, we shall consider a Dirac particle described by the Hamiltonian given by Eq. (\ref{Dirac_eq}) subject to a double well one-dimensional potential $U\left(x\right)$. Hereafter, the valley index number, $s_{\mathrm{K}}$, is set to one. The other valley's wave function can be readily obtained by performing a  sign change on  $k_{y}$. The Hamiltonian acts on the two-component Dirac wavefunction
\begin{equation}
\Psi=\left(\begin{array}{c}
\psi_{A}\left(x\right)\\
\psi_{B}\left(x\right)
\end{array}\right)e^{ik_{y}y}
\label{eq:wave_start}
\end{equation}
to yield the coupled first-order differential equations 
\begin{equation}
\left(V-E+\Delta_{z}\right)\psi_{A}-i\left(\frac{d}{d\tilde{x}}+\Delta_{y}\right)\psi_{B}=0\label{eq:origin_1}
\end{equation}
and 
\begin{equation}
\left(V-E-\Delta_{z}\right)\psi_{B}-i\left(\frac{d}{d\tilde{x}}-\Delta_{y}\right)\psi_{A}=0,\label{eq:origin_2}
\end{equation}
where $\tilde{x}=x/L$ and $L$ is a constant with the dimension of length. $V=UL/\hbar v_{\mathrm{F}}$
and the charge carrier energy, $\varepsilon$, have been scaled such
that $E=\varepsilon L/\hbar v_{\mathrm{F}}$. The charge carriers propagate along the y-direction with wave vector $k_{y}=\Delta_{y}/L$, which is measured relative to the Dirac point and $\Delta_{z}= k_{z}L$ represents their effective mass in dimensionless units. Finally, $\Psi_{A}\left(x\right)$ and $\Psi_{B}\left(x\right)$ are the wavefunctions associated with the $A$ and $B$ sub-lattices respectively. Substituting 
\begin{equation}
\psi_{A}=\frac{1}{2}\left[\psi_{1}\exp\left(\frac{1}{2}i\phi\right)+\psi_{2}\exp\left(-\frac{1}{2}i\phi\right)\right]
\nonumber
\end{equation}
and 
\begin{equation}
\psi_{B}=\frac{1}{2}\left[\psi_{1}\exp\left(\frac{1}{2}i\phi\right)-\psi_{2}\exp\left(-\frac{1}{2}i\phi\right)\right],
\nonumber
\end{equation}
where $\phi=\arctan\left(\Delta_{y}/\Delta_{z}\right)$, allows Eqs.(\ref{eq:origin_1},\ref{eq:origin_2})
to be reduced to a single second-order differential equation in $\psi_{j}$: 
\begin{equation}
-\frac{d^{2}\psi_{j}}{dZ^{2}}+\mathrm{V}_{s_{j}}\psi_{j}=M^{2}\psi_{j},
\label{eq:super_sym_1}
\end{equation}
where
\begin{equation}
\mathrm{V}_{s_{j}}\left(Z\right)=W^{2}\left(Z\right)-s_{c}s_{j}\frac{dW\left(Z\right)}{dZ}
\label{eq:non_linear}
\end{equation}
and the other spinor component is found by the relation
\begin{equation}
\psi_{j'}=-\frac{1}{M}\left(V-E+s_{c}s_{j}\frac{d}{dZ}\right)\psi_{j},
\end{equation}
where $\tilde{x}=s_{c}iZ$, $s_{c}=\pm1,$ $W=V-E$, $s_{j}=\left(-1\right)^{j}$,
and $j=1,\,2$ ($j'=2,1$) corresponds to the spinor components $\psi_{1}$ ($\psi_{2}$) and
$\psi_{2}$ ($\psi_{1}$)  respectively. In this basis the particle's transverse momentum $\Delta_{y}$ and in-plane effective mass $\Delta_{z}$, have been combined into a single effective mass, $M=\sqrt{\Delta_{y}^{2}+\Delta_{z}^{2}}$. Eq.~(\ref{eq:super_sym_1}) is of same form as the Schrodinger equation, i.e., a second order differential equation with no first order derivative. Therefore, if $\mathrm{V}_{s_{j}}$ is equal to a potential possessing known solutions to the Schr\"{o}dinger equation, then for zero energy states, we can readily write down the bound state spectrum of the potential, $W$, which  satisfies the non-linear Eq.~(\ref{eq:non_linear})~\cite{dutt1988supersymmetry,cooper1988supersymmetry,ho2014zero}. It should also be noted that for zero energy modes, the left-hand side of Eq.~(\ref{eq:super_sym_1}) is of the form of the super-Hamiltonian, $W$ plays the role of the superpotential and the allowed $M$ plays the role of the energy eigenstates~\cite{dutt1988supersymmetry,cooper1988supersymmetry}. Thus there exists a plethora of potentials which admit solutions for zero-energy modes. For non-zero energy, the $W$ which satisfies the non-linear Eq.~(\ref{eq:non_linear})
for a given potential $\mathrm{V}_{s_{j}}$ will be eigenvalues of an energy-dependent potential \cite{schulze2017bound}. However, we are interested in potentials which are independent of energy and are suitable for the use of modelling double-wells in Dirac materials. 

The solutions to many second order differential equations, such as the Schr\"{o}dinger equation and the 2D Dirac equation, can be obtained in terms of hypergeometric functions~\cite{landau2013quantum,cook1971relativistic, Moshinsky_1989, kennedy2002woods, guo2009transmission, candemir2013massive, ishkhanyan2018schrod}. Indeed any second order differential equation, possessing three regular singularities, can be re-expressed as Euler’s Hypergeometric differential equation. Terminating the resulting hypergeometric series, and/or utilizing their well-known connection formulae allows the eigenvalues of many potentials to be obtained. However, when considering the 2D Dirac equation for double-well potentials, the resulting second order differential equation's may contain more than three regular singularities. For example, four regular singularities means the differential equation can be re-expressed as Heun’s Equation~\cite{heun1888theorie}. For the family of potentials which result in a differential equation having two regular singularities and one irregular singularity of rank 1, the second order differential equation can be transformed into the confluent form of Heun’s equation. Indeed, reducing a system of coupled first-order differential equations to the confluent Heun equation has been exploited to solve various generalisations of the quantum Rabi model~\cite{kocc2002quasi,xie2017quantum}. Despite the absence of a general connection formula, the confluent Heun functions can still serve as a powerful tool in studying confinement potentials, and has been extensively applied in the fields of General Relativity and Quantum Gravity~\cite{hortaccsu2018heun}.

In what follows, we consider the case where the 2D Dirac equation reduces to the confluent Heun equation for a family of potentials, some of which can be used to describe a double-well separated by a barrier. This quantum model is quasi-exactly solvable~\cite{turbiner1988quantum,ushveridze2017quasi,bender1998quasi,kocc2002quasi,downing2013solution,hartmann2014quasi,hartmann2014bound,hartmann2017two}, i.e. only a subset of the eigenvalues can be found explicitly. We study bound states contained within double-well potentials
and calculate their entire energy spectrum. Energy level splitting associated with quantum tunneling between wells, and avoided crossings associated with the inter-mixing of electron-hole states are discussed. Finally, terahertz (THz) applications utilizing the doublet states and avoided crossing points are considered.

Let us search for transformations which may be performed on the dependent and independent variables of Eq.~(\ref{eq:super_sym_1}),
and the corresponding energy-independent potentials which allow Eq.~(\ref{eq:non_linear}) to be transformed into the confluent Heun equation. In some instances, the resulting confluent Heun series can be terminated~\cite{ronveaux1995heun}, allowing a subset of the eigenvalues to be obtained exactly. In other instances, the entire energy spectrum can be obtained fully via the Wronskian method~\cite{zhong2013analytical,maciejewski2014full,hartmann2014bound,hartmann2014quasi,xie2017quantum,hartmann2017two}. Similar approaches have been implemented in the non-relativistic regime, indeed, for the Schr\"{o}dinger equation there exists 35 choices for the coordinate transformation, each leading to eleven independent potentials which are exactly solvable in terms of the general Heun functions~\cite{ishkhanyan2015thirty,ishkhanyan2018schrod}. The Schr\"{o}dinger equation also reduces to various forms of confluent Heun equations for potentials such as the Natanzon family~\cite{ishkhanyan2016schrodinger} and several others~\cite{ishkhanyan2016discretization}.

The B$\mathrm{\hat{o}}$cher symmetrical form of the Confluent Heun equation, also known as the generalized spheriodal equation, can be
written using the notation from Ref.~\cite{ronveaux1995heun}
\begin{equation}
\frac{d}{d\xi}\left[\left(\xi^{2}-1\right)\frac{dy\left(\xi\right)}{d\xi}\right]+\left[-p^{2}\left(\xi^{2}-1\right)+2p\beta\xi-\lambda-\frac{m^{2}+s^{2}+2ms\xi}{\xi^{2}-1}\right]y\left(\xi\right)=0,
\label{eq:CHE_y}
\end{equation}
where the regular singular points are located at $\xi=1$ and $\xi=-1$. The first order derivative can be removed by transforming the independent
variable to $v\left(\xi\right)=y\left(\xi\right)\sqrt{1-\xi^{2}}$,
this allows Eq.~(\ref{eq:CHE_y}) to be written in normal symmetrical form:
\begin{equation}
-\frac{d^{2}v\left(\xi\right)}{d\xi^{2}}+\left[p^{2}+\frac{\lambda-2p\beta\xi}{\xi^{2}-1}+\frac{m^{2}+s^{2}-1+2ms\xi}{\left(\xi^{2}-1\right)^{2}}\right]v\left(\xi\right)=0.
\label{eq:poly_form}
\end{equation}
We shall now re-express Eq.~(\ref{eq:super_sym_1}), into the same form as Eq.~(\ref{eq:poly_form}). There are many transformations which may be performed on the dependent and independent variables of Eq.~(\ref{eq:super_sym_1}),
which give rise to a second order differential equation possessing no first order derivative. Applying the transformation of the independent
variable $\xi=\xi\left(Z\right)$ and $\psi_{j}=\exp\left[-\int\frac{1}{2}\left(\frac{1}{\Phi}\frac{d\Phi}{d\xi}\right)d\xi\right]\chi_{j}$
to the dependent variable yields

\begin{equation}
-\frac{d^{2}\chi_{j}}{d\xi^{2}}+\left[\frac{1}{4}\left(\frac{1}{\Phi}\frac{d\Phi}{d\xi}\right)^{2}+\frac{1}{2}\frac{d}{d\xi}\left(\frac{1}{\Phi}\frac{d\Phi}{d\xi}\right)-s_{c}s_{j}\frac{1}{\Phi}\frac{dV}{d\xi}+\frac{\left(V-E\right)^{2}-M^{2}}{\Phi^{2}}\right]\chi_{j}=0
\end{equation}
where $\Phi=\frac{d\xi\left(Z\right)}{dZ}$. The potential
\begin{equation}
V=\frac{a_{2}\xi^{2}+a_{1}\xi+a_{0}}{1-\xi^{2}}\Phi_{\eta_{1},\,\eta_{2}}
\label{eq:Gen_Potential}
\end{equation}
represents a family of energy-independent potentials which for the
quasi-one-dimensional Dirac problem admits wavefunctions in terms
of the confluent Heun functions. Here $\Phi_{\eta_{1},\,\eta_{2}}=C_{\eta_{1},\,\eta_{2}}\left(1-\xi\right)^{\eta_{1}}\left(1+\xi\right)^{\eta_{2}}$,
$\eta_{1}$ and $\eta_{2}$ can take the values of 1 and 0, and
$C_{\eta_{1},\,\eta_{2}}$ is a constant. As mentioned in the introduction, this family of potentials belongs to the class of quantum models which are quasi-exactly solvable~\cite{turbiner1988quantum,ushveridze2017quasi,bender1998quasi,kocc2002quasi,downing2013solution,hartmann2014quasi,hartmann2014bound,hartmann2017two}, where only some of the eigenvalues can be expressed explicitly. 

When $\eta_{1}=\eta_{2}=1$, Eq.~(\ref{eq:Gen_Potential}) can be expressed as the Rosen-Morse potential, $V=b_{1}\tanh^{2}\left(\tilde{x}\right)+b_{2}\tanh\left(\tilde{x}\right)$, whose eigenvalue spectrum, which is quasi-exact, was discussed in ref.~\cite{hartmann2017two}. The
analytic bound state energy spectra of the Rosen-Morse Potential for the two-dimensional Dirac problem has also been obtained via the Nikiforov-Uvarov
method~\cite{ikhdair2010approximate}. When $\eta_{1}\neq\eta_{2}$, Eq.~(\ref{eq:Gen_Potential}) can be
expressed as a generalized Hulthen-like potential:
\begin{equation}
V=\frac{c_{1}\exp\left(-2\lambda Z\right)+c_{2}\exp\left(-\lambda Z\right)+c_{3}}{1+c_{4}\exp\left(-\lambda Z\right)},\label{eq:Hulthen}
\end{equation}
where $\lambda$ and $c_{1,2,3,4}$ are consants. When $c_{1}=0$
and $c_{4}=-1$, Eq.~(\ref{eq:Hulthen}) becomes a linear combination of the Hulthen potential~\cite{Hulthen1942} and its logarithmic derivative. The Hulthen
potential has previously been investigated for the 2-D Dirac equation using an algebraic approach~\cite{roy1990dirac}. The case where~Eq.~(\ref{eq:Hulthen}) reduces to a single exponential
has also been studied in graphene waveguides ~\cite{stone2012searching}. It should also be noted that when $\Phi=i$, the potential given by Eq.~(\ref{eq:Gen_Potential}) can be reduced to the shifted 1D Coulomb potential, which has applications to charged impurities and excitons in carbon nanotubes~\cite{downing2014one}. 

Let us return to the more general form of the potential given by Eq.~(\ref{eq:Gen_Potential}), one of the opportunities to describe a smooth double well is to choose the case of $\eta_{1}=\eta_{2}=0$.  We require that our potential is non-singular and vanishes as $x\rightarrow\pm\infty$.
Therefore we set, $a_{2}=0$ and the potential becomes
\begin{equation}
V=\frac{a_{1}Z+a_{0}}{1-Z^{2}}\equiv\frac{A_1\tilde{x}+A_{0}}{1+\tilde{x}^{2}},
\label{eq:pot_tobe_solved}
\end{equation}
where the potential parameters $A_{1}$ and $A_{0}$ are related to the parameters appearing in Eq.~(\ref{eq:poly_form})
via the relations:
\begin{equation}
\nonumber
p=s_{p}\sqrt{E^{2}-M^{2}},\qquad\beta=-is_{c}A_1E/p,
\end{equation}
\begin{equation}
\nonumber
\lambda=-A_1\left(A_1+is_{j}\right)+2EA_{0},
\end{equation}
\begin{equation}
\Lambda_{1,\,2}=s_{a}A_{0},\qquad\Lambda_{2,\,1}=is_{a}\left(A_1+is_{j}\right)s_{c},
\nonumber
\end{equation}
where $\Lambda_{1}=m$, $\Lambda_{2}=s$ and $s_{a,\,p}=\pm1$.  Equation~(\ref{eq:pot_tobe_solved})
is a linear combination of the Lorentzian and its logarithic derivative, which has known solutions for the Radial Dirac equation \cite{downing2011zero}. 

Since the potential given by Eq.~(\ref{eq:pot_tobe_solved}) is smooth and vanishes as $x\rightarrow\pm\infty$, it may be utilized for modelling top-gated structures in 2D Dirac materials. The Lorentzian has already been used to model the potential generated by a top-gate formed by a carbon nanotube~\cite{cheng2019guiding}. It should also be noted that when $p=0$, i.e. $\left| E\right|= M$, the Lorentzian admits exact energy eigenvalues (see Appendix \ref{sec:Lorentzian}). Exponentially decaying potentials also play an important role in the modelling of heterostructure-devices based upon zero-energy modes in Dirac-materials, as these potentials are often quasi-exact, admitting some exact energy eigenvalues. 
As mentioned previously in the introduction, the majority of work on multiple-barrier structures focus on smooth but sharp potentials.  However, realistic potential profiles vary slowly over the length scale of the Dirac material's lattice constant, and discontinuous potentials have yet to be realized. Furthermore, many piecewise potentials do not result in smooth wavefunctions across the whole of configuration space due to the nontrivial nature of their boundary conditions~\cite{xu2015guided,xu2016guided,you2017modes}. Smooth potentials do not suffer from this problem ~\cite{hartmann2010smooth, hartmann2014quasi, hartmann2017two, hartmann2017pair}, additionally they permit inter-valley scattering to be neglected.

\begin{figure}
  \includegraphics[width=0.9\linewidth]{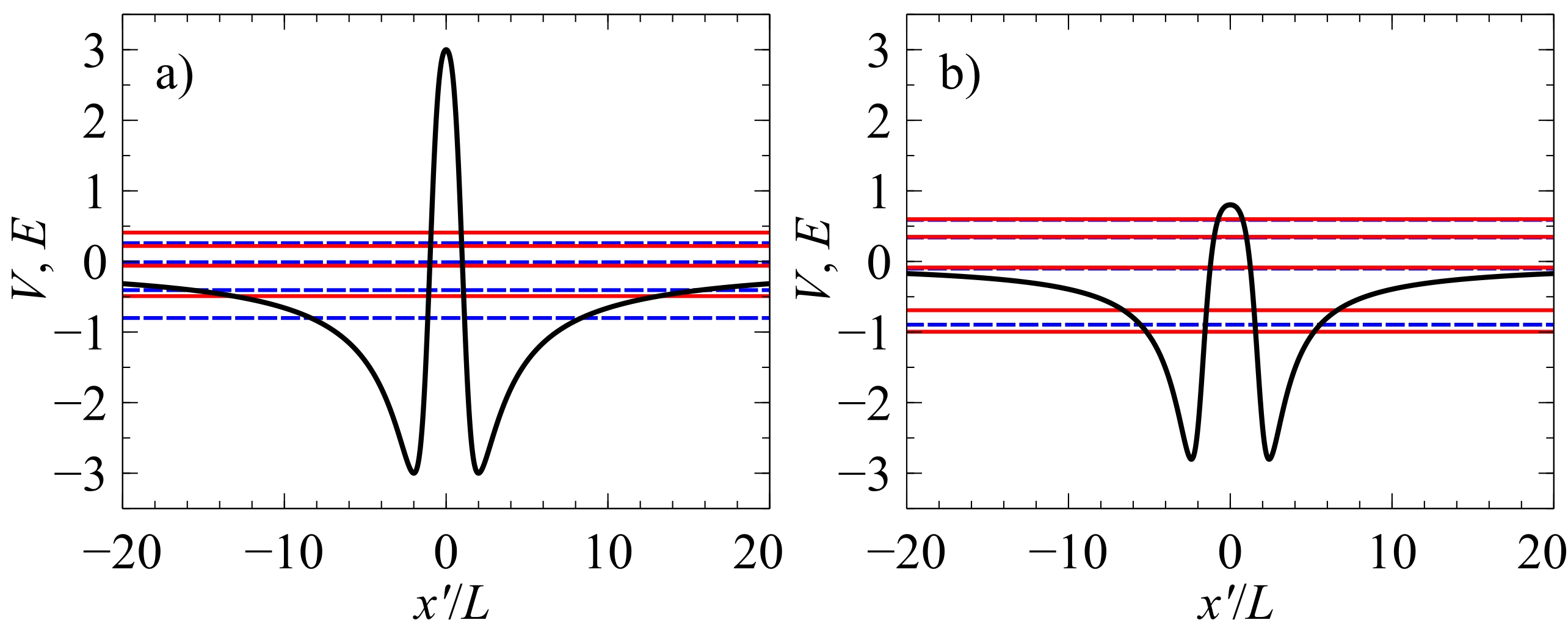}
  \caption{Energy levels of 2D Dirac electrons in a double-well potential. The black solid lines show the two potentials given by Eq. (\ref{eq:pot_tobe_solved}), (a) for the case of $A_{1} = -6$ and $A_{0} = 0$ and (b) for the case of $A_{1} = -3$ and $A_{0} = -2$. The horizontal lines depict the bound state energies for the four lowest doublet states, plus all the hole-like states in the spectrum for the case of $M=1$. The red solid and blue dashed lines correspond to the eigenvalues of the even and odd modes of $\psi_I$ respectively.
  }
\label{fig:V_Plot}
\end{figure}

The stationary points,
$\tilde{x}_{s_{r}}$, of the potential Eq.~(\ref{eq:pot_tobe_solved})
are located at 
\begin{equation}
\tilde{x}_{s_{r}}=-r+s_{r}\sqrt{1+r^{2}},
\end{equation}
where $r=A_{0}/A_1$ and $s_{r}=\pm1$. Making the coordinate transformation $\tilde{x}=\left|x'\right|+\tilde{x}_{s_{r}}$, allows the potential Eq.~(\ref{eq:pot_tobe_solved}) to be utilized as a model for quasi-one-dimensional double-wells in realistic top-gated Dirac material heterostructures. When $A_1<0$ and $s_r=-1$ the potential is a double well (see Fig.~\ref{fig:V_Plot}), whose local minima are located a distance of $2\sqrt{1+r^{2}}$ apart. The two wells, of depth $A_{1}/(2\tilde{x}_{1})$ are separated by a barrier, of height $A_{1}/(2\tilde{x}_{-1})$, which acts as a single-well for hole-like particles, with the possible experimental set-up shown in Fig.~(\ref{fig:my_label}). Although we are dealing with a single-particle Hamiltonian, it is convenient to call the states with the energy growing with increasing $|\Delta_{y}|$ when $|\Delta_{y}|\rightarrow\infty$ as electron-like states or electrons, whereas the hole-like states correspond to $E\rightarrow-\infty$ when $|\Delta_{y}|\rightarrow\infty$.
In the limit that $r\gg1$, the central barrier's height becomes negligible in comparison to the depth of the two wells. Conversely, when $A_1>0$ and $s_r=-1$, the potential becomes a single electron well, lying between two barriers, which act as a double well for hole-like particles. 

The non-symmetrical canonical form of the confluent Heun equation can be written as
\begin{equation}
\frac{\partial^{2}u\left(\tilde{\xi}\right)}{\partial\tilde{\xi}^{2}}+\left(4p+\frac{\gamma}{\tilde{\xi}}+\frac{\delta}{\tilde{\xi}-1}\right)\frac{\partial u\left(\tilde{\xi}\right)}{\partial\tilde{\xi}}+\frac{4p\alpha\tilde{\xi}-\sigma}{\tilde{\xi}\left(\tilde{\xi}-1\right)}u\left(\tilde{\xi}\right)=0
\label{eq:CHE_non_sym}
\end{equation}
where 
\begin{equation}
\tilde{\xi}=\frac{1-\xi}{2},\qquad\gamma=m+s+1,\qquad\delta=m-s+1,
\nonumber
\end{equation}
\begin{equation}
\alpha=-\beta+m+1,\qquad\sigma=\lambda-2p\left(\beta-\gamma\right)-m\left(m+1\right)
\nonumber
\end{equation}
and $u=u\left(p,\,\alpha,\,\gamma,\,\delta,\,\sigma;\,\frac{1-\xi}{2}\right)$ are the Heun Confluent functions~\cite{ronveaux1995heun}, 
which are related to the solutions of $v\left(\xi\right)$ of Eq.~(\ref{eq:poly_form}) by the relation:
\begin{equation}
v\left(\xi\right)=\left(1-\xi\right)^{\frac{\gamma}{2}}\left(1+\xi\right)^{\frac{\delta}{2}}e^{-p\xi}u\left(\frac{1-\xi}{2}\right),
\label{eq:link}
\end{equation}
which from hereon we shall denote as $v\left(\xi\right)=\Psi_{s_{j},s_{c}}$. Currently, there are no universal expressions for arbitrary parameters connecting Heun functions about different singular points, and unlike for Gauss hypergeometric functions, there are no expressions relating the derivative of a confluent Heun function to another confluent Heun function, although particular instances have been obtained~\cite{fiziev2009novel,shahnazaryan2014new,hartmann2017two}. The dearth of connection formulae makes obtaining the analytic expressions for the complete eigenvalue spectrum of a wavefunction expressed in terms of confluent Heun function  non-trival. However, for the potential, Eq.~(\ref{eq:pot_tobe_solved}), 
the derivative can be expressed exactly in terms of Heun Confluent function with new parameters. Using the definition of the Heun Confluent function~\cite{ronveaux1995heun}, $\Psi_{s_{j},s_{c}}$ can be expressed in terms $\Psi_{-s_{j},s_{c}}$ via the relation:
\begin{equation}
-\frac{1}{M}\left[\left(V-E\right)\Psi_{s_{j},s_{c}}+s_{j}i\frac{d\Psi_{s_{j},s_{c}}}{dx}\right]=
\left(-\frac{s_{a}+A_{0}+is_{c}A_1}{2M}\right)^{-s_{j}s_{c}s_{a}}\Psi_{-s_{j},s_{c}},
\label{eq:Heun_Identity}
\end{equation}
which allows the solution to Eq.~(\ref{eq:non_linear}) for $\psi_{1}$ and its corresponding $\psi_{2}$ to be be written as
\begin{equation}
\psi_{1}=\sum_{s_{c},c_{p},s_{a}}C_{s_{c},c_{p},s_{a}}\left(-\frac{s_{a}+A_{0}+is_{c}A_1}{2M}\right)^{-\frac{s_{c}s_{a}}{2}}\Psi_{-1,s_{c}}
\label{eq:new_fun_nonsym1}
\end{equation}
and
\begin{equation}
\psi_{2}=\sum_{s_{c},c_{p},s_{a}}C_{s_{c},c_{p},s_{a}}\left(-\frac{s_{a}+A_{0}+is_{c}A_1}{2M}\right)^{\frac{s_{c}s_{a}}{2}}\Psi_{1,s_{c}}
\label{eq:new_fun_nonsym2}
\end{equation}
respectively, where $m=s_{a}A_{0}$, $s=is_{a}\left(A_1+is_{j}\right)s_{c}$, 
and $C_{s_c,c_p,s_a}$ are weighting coefficients. It should be noted that if $m$ and $s$ are exchanged, then the phase factor appearing in Eq.~(\ref{eq:Heun_Identity}) and Eq.~(\ref{eq:new_fun_nonsym2}) must be multiplied by the factor $4^{s_{c}s_{a}}$. For bound states, we require that the term $e^{-p\xi}$ decays as $x\rightarrow \infty$, this imposes two conditions: First $s_p=s_c$, and second, the absolute value of the particle's energy must be less than the reduced mass, i.e., $\left|E\right|<\left|M\right|$.

The functions $\psi_{1}$ and $\psi_{2}$ are neither even nor odd. For understanding optical selection rules and for a more clear mapping of our results to the conventional double well picture we shall move to the symmterized basis functions $\left|\psi_{I}\right\rangle =\left|\psi_{1}\right\rangle -\left|\psi_{2}\right\rangle$ and  $\left|\psi_{II}\right\rangle =-i\left(\left|\psi_{1}\right\rangle +\left|\psi_{2}\right\rangle \right)$. It should be noted that in this basis, an exchange of the sign of $V$ and $E$ is formally equivalent to exchanging $\psi_I$ with $\psi_{II}$ for the original $V$ and $E$. Therefore, inverting the potential only results in a sign change of the eigenvalues. 
In the symmetrized basis one can construct a linear combination of functions Eq.~(\ref{eq:new_fun_nonsym1}) and Eq.~(\ref{eq:new_fun_nonsym2}) to form solutions which are electron- ($s_a=1$) and hole-like ($s_a=-1$):
\begin{equation}
\psi_{I}=\sum_{s_{a}}D_{s_{a}}\Im\left(\rho^{\star}\Psi_{1,1}+\rho^{-1}\Psi_{1,-1}\right)
\label{eq:elecron_hole_wave1}
\end{equation}
\begin{equation}
\psi_{II}=\sum_{s_{a}}D_{s_{a}}\Re\left(\rho^{\star}\Psi_{1,1}+\rho^{-1}\Psi_{1,-1}\right)
\label{eq:elecron_hole_wave2}
\end{equation}
where $\Re$ and $\Im$ are the standard notation for the real and imaginary parts of a complex number, $\rho=\left(-2M\right)^{-\frac{s_{a}}{2}}\left(s_{a}+A_{0}-iA_1\right)^{\frac{s_{a}}{2}}$  and $D_{s_{a}}$ are the weighting constants. For bound states we require that the spinor components, Eq.~(\ref{eq:elecron_hole_wave1}) and Eq.~(\ref{eq:elecron_hole_wave2}), vanish at infinity, i.e., $\psi_{I,II}(x'\longrightarrow \pm \infty)=0$. While the coordinate transformation,  $\tilde{x}=\left|x'\right|+\tilde{x}_{s_{r}}$, imposes the continuity condition
\begin{equation}
\psi_{I}\left(\tilde{x}_{-}\right)=0,
\label{eq:parity_1}
\end{equation}
for odd-states and
\begin{equation}
\left.\frac{\partial\psi_{I}}{\partial x}\right|_{\tilde{x}_{-}}=0,
\label{eq:parity_2}
\end{equation}
for even-states. 
The coordinate transformation also requires that $\psi_{I,II}\left(x'\right)=-\psi_{I,II}\left(-x'\right)$ and $\psi_{I,II}\left(x'\right)=\psi_{I,II}\left(-x'\right)$ for odd and even states respectively.
However, the radius of convergence, of the power series, 
is unity. Therefore, an iterative, analytic continuation method must be employed to evaluate the confluent Heun function beyond its radius of convergence~\cite{motygin2018evaluation}. First  the  power series, 
$u$ is evaluated at the point $\widetilde{\xi}_1$, which lies within the radius of convergence. A new series expansion is then performed about the point $\widetilde{\xi}_1$, which in turn is used to evaluate the point $\widetilde{\xi}_2$, and so on and so forth. This allows the energy eigenvalues to be found via a simple shooting method \cite{press1992numerical,killingbeck1987shooting} which utilizes the wave-functions of Eq.~(\ref{eq:new_fun_nonsym1}) and Eq.~(\ref{eq:new_fun_nonsym2}).

\begin{figure}[ht]
\includegraphics[width=0.25\linewidth]{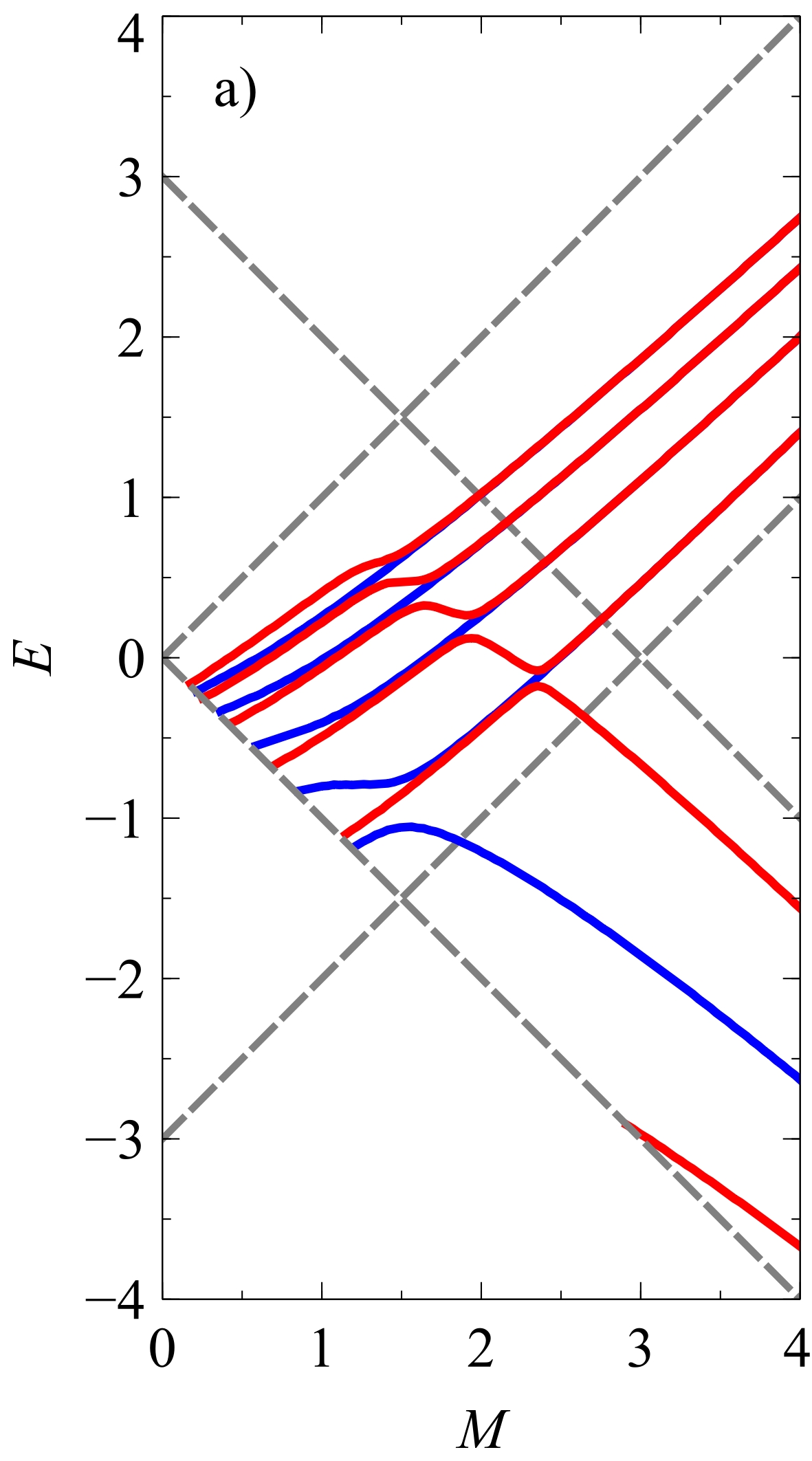}
\includegraphics[width=0.25\linewidth]{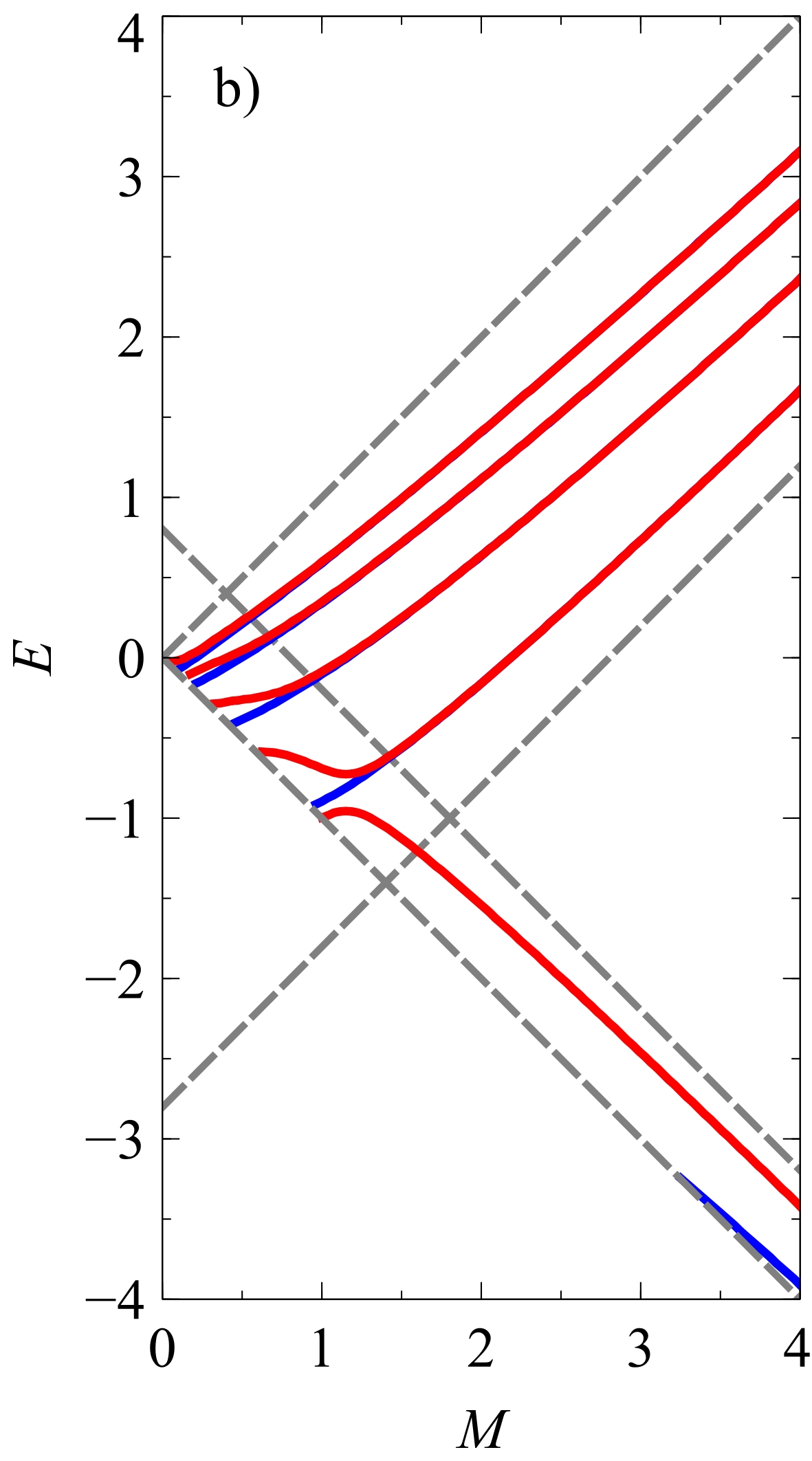}
\caption{
The energy spectrum of bound states contained within a double-well separated by a barrier, defined by the potential parameters (a) $A_{1}=-6$ and $A_{0}=0$ and (b) $A_{1}=-3$ and $A_{0}=-2$, as a function of effective mass, $M$. Here, only the four lowest doublet states are displayed, and all barrier states are present within the energy range shown. The alternating red and blue lines represent the even (odd) and odd (even) modes of $\psi_I$ $(\psi_{II})$ respectively. 
The boundary at which the bound states merge with the continuum is denoted by the grey-dashed lines.
}
\label{fig:Disp_Old}
\end{figure}

\begin{figure}
  \includegraphics[width=0.90\linewidth]{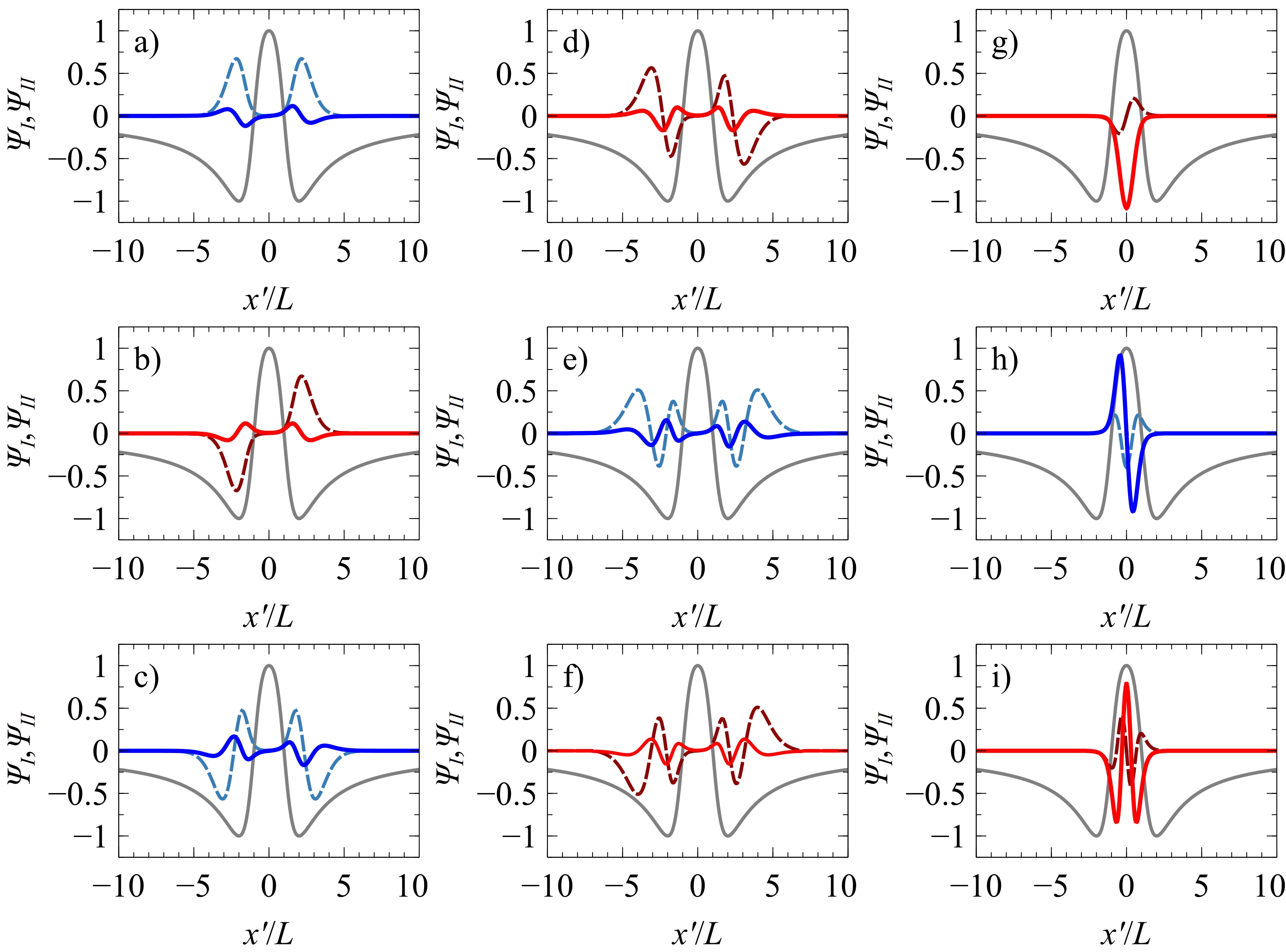}
\caption{
The normalized bound-state wavefunctions of the double-well potential. $V=A_{1}\tilde{x}/(1+\tilde{x}^2)$, where $\tilde{x}=|x'|-1$, $A_{1}=-6$ and $M=3.5$ for the first 3 doublet states of energies: 
(a) $E = 0.92814$,
(b) $E = 0.92818$,
(c) $E = 1.54719$,
(d) $E = 1.54723$,
(e) $E = 1.98230$,
(f) $E = 1.98235$,
and for the hole states of energy (g) $E = -1.10843$, (h) $E = -2.23465$ and (i) $E = -3.31137$. The solid red and blue lines correspond to the even and odd modes of $\Psi_I$ respectively. While the dashed lines corresponds to the other spinor component $\Psi_{II}$. The grey line shows the double-well potential as a guide to the eye.
}
\label{fig:wave_1}
\end{figure}

In Fig.~(\ref{fig:Disp_Old}) we plot the energy spectrum  of bound states (square-integrable along the x-direction) defined by the potential parameters $A_{1}=-6$ and $A_{0}=0$,  as a function of effective mass, $M$, displaying only the four lowest lying electron-like doublets and all the hole-like states within the energy range.
When $E \ll M+V_{min}$, (where $V_{min}$ is the minimum value of the potential)
the spectrum resembles a single, hole-like particle, trapped within a positive potential barrier. The highest in energy hole-like branch is $s$-like (nodeless) for $\psi_I$, and the number of nodes increases by one as the branches progress towards more negative energies. 
For $E \gg -M+V_{max}$, (where $V_{max}$ is the maximum value of the potential), the spectrum is electron-like, and the energy splittings of the doublets (each shown by a red line with a blue line underneath in Fig.~(\ref{fig:Disp_Old})) tend to zero with increasing $M$, as expected for non-relativistic particles when increasing mass suppresses tunneling. In this regime, it can be seen from Fig.~(\ref{fig:wave_1}) that the particles are localized in the region of the wells, and that the wavefunctions behave as a linear combination of the wavefunctions associated with the individual wells. This gives rise to two states where there would of been one if tunneling was forbidden. For $\psi_I$, the number of nodes increases by one as the branches progress to higher energies. However, since $\psi_{II}$ is related to the derivative of $\psi_I$ and the potential is even in $x$, i.e. $V(-x)=V(x)$, the asymmetric linear combination of the two single well functions forces a node in the barrier between them, while the symmetric combination does not. Which is reflected in the behavior of $\psi_{II}$ being precisely the same as a wavefuntion for a non-relativistic particle in a double well, whereas, the component $\psi_{I}$ has the opposite parity. Finally, in the energy zone $M-V_{min}<E<-M+V_{max}$, the solutions are a linear combination of electron-like and hole-like states. It can be seen from see Fig.~(\ref{fig:Disp_Old}) that a series of avoided crossings are opened in the energy spectrum due to the repulsion of barrier and well-doublet states. 
Also in this  energy zone, the energy level splitting associated with quantum tunneling increases, until the bound states merge with the continuum states. This energy splitting between the two doublets can be controlled in realistic Dirac-material-heterostructures by either, adjusting the strength of the applied top-gate voltages, or, by the choice of geometry. This, coupled with the ability to change the position of the Fermi level via a back gate, gives rise to many possible device applications. Indeed, utilizing the dependence on the number of zero-energy modes on potential strength in single-well smooth electron waveguides, has been proposed as the basis switching devices in graphene~\cite{hartmann2010smooth}. 

\section{Terahertz Transitions}\label{Transitions}
Within this section we provide the general formalism for calculating the dipole matrix element of THz transitions between guided modes of quasi-particles described by the Hamiltonian given by Eq.~(\ref{Dirac_eq}). For a Dirac particle subject to a 1D potential $U(x)$, the unperturbed Hamiltonian is given by $\hat{H}_0=\hat{H} +  \mathrm{\boldsymbol{I}}U_{z}$,
and the corresponding eigenfunctions of the unperturbed Hamiltonian are given by $\left(\psi_{A}\left(x\right),\psi_{B}\left(x\right)\right)^{T}e^{ik_{y}y}/\sqrt{N}$, 
where $N$ is a normalization factor given by the expression, $N=l\intop_{-\infty}^{\infty}\left(\left|\psi_{A}\right|^{2}+\left|\psi_{B}\right|^{2}\right)dx$, and $l$ is the sample length. In the presence of an electromagnetic field, the particle momentum operator, $\hat{\boldsymbol{p}}$, is modified such that $\hat{\boldsymbol{p}}\rightarrow\hat{\boldsymbol{p}}+e\boldsymbol{A}/c$, where $e$ is the elementary charge, and $\boldsymbol{A}$ is the magnetic vector potential, which is related to $\hat{\boldsymbol{e}}=\left(e_{x},e_{y}\right)$, the unit vector describing the polarization of the electromagnetic wave, via the relation $\boldsymbol{A}=A\hat{\boldsymbol{e}}$. For linearly polarized light, the polarization vector is expressed as $\left(\cos\left(\varphi_{0}\right),\sin\left(\varphi_{0}\right)\right)
$, while for right- and left-handed polarized light it is
$\left(1,-i\right)/\sqrt{2}$ and $\left(1,i\right)/\sqrt{2}$, respectively. The general form of the perturbation due to an electromagnetic wave impinging normally to a Dirac material is
\begin{equation}
\delta H=
\frac{eA v_{\mathrm{F}}}{c}\left(\sigma_{x}e_{x}+s_{\mathrm{K}}\sigma_{y}e_{y}\right).
\end{equation}
Using the wavefunctions given in Eq.~(\ref{eq:wave_start}), in the limit that $l\longrightarrow\infty$ the matrix element of 
transition becomes:
\begin{equation}
\left|\left\langle f\left|\delta H\right|i\right\rangle \right|=
G_{1}
\left|\intop_{-\infty}^{\infty}\left[\left(e_{x}-is_{\mathrm{K}}e_{y}\right)\psi_{A,f}^{\star}\psi_{B,i}+\left(e_{x}+is_{\mathrm{K}}e_{y}\right)\psi_{B,f}^{\star}\psi_{A,i}\right]dx\right|
\delta_{k_{y,i},k_{y,f}},
\label{eq:optical_matrix_element}
\end{equation}
where $G_{1}=eA v_{\mathrm{F}}/\left(c \sqrt{N_{i}N_{f}}\right)$,
and the indices $i$ and $f$ correspond to the initial and final states respectively. For free two-dimensional massless Dirac fermions, Eq.~(\ref{eq:optical_matrix_element}) becomes valley independent~\cite{saroka2018momentum}. While in contrast, massive two-dimensional Dirac fermions have valley-dependent optical transition rules~\cite{hartmann2019interband}. We shall now analyze the optical selection rules between guided modes contained within the electrostatically-controlled double well defined by Eq.~(\ref{eq:pot_tobe_solved}). To do so it is more convenient to move from the original $\left|\psi_{A}\right\rangle$, $\left|\psi_{B}\right\rangle$ basis to the symmetrized one. The functions $\psi_{A}$ and $\psi_{B}$ can be expressed in terms $\psi_{I}$ and $\psi_{II}$ via the transformation $\left(\psi_{A},\psi_{B}\right)^{T}=U\left(\psi_{I},\psi_{II}\right)^{T}$, where
\begin{equation}
U=\frac{1}{2}\left(\begin{array}{cc}
i\sin\left(\frac{1}{2}\phi\right) & i\cos\left(\frac{1}{2}\phi\right)\\
\cos\left(\frac{1}{2}\phi\right) & -\sin\left(\frac{1}{2}\phi\right)
\end{array}\right).
\end{equation}
To shift to the symmterized basis we make a unitary transformation to $\delta H$ with the unitary operator $U$. Under this change, the perturbation $\delta H$ transforms as $\delta\breve{H}=U^{\dagger}\delta H U$:
\begin{equation}
\delta\breve{H}=-\frac{e v_{\mathrm{F}}A}{4c}\left(e_{x}\sigma_{y}+\frac{s_{\mathrm{K}}e_{y}\Delta_{z}}{\sqrt{\Delta_{z}^{2}+\Delta_{y}^{2}}}\sigma_{x}+\frac{e_{y}\Delta_{y}}{\sqrt{\Delta_{z}^{2}+\Delta_{y}^{2}}}\sigma_{z}\right)
\end{equation}
The optical transitions between different guided modes can be categorized into two distinct groups. First, when the parity of each of the symmetrized spinor components $\psi_{I}$ and $\psi_{II}$, is preserved during the transition, i.e., both $\psi_{I,i}$ and $\psi_{I,f}$ are even, or both odd. Second, when the parity of each component changes after the transition, i.e., $\psi_{I,i}$ is odd (even), while $\psi_{I,f}$ is even (odd). When a transition, which preserves the parity of each spinor component, occurs, its matrix element given by Eq.~(\ref{eq:optical_matrix_element}), can be expressed in terms of the symmetrized spinor components, Eq.~(\ref{eq:elecron_hole_wave1}) and Eq.~(\ref{eq:elecron_hole_wave2}) as
\begin{equation}
\left|\left\langle f\left|\delta H\right|i\right\rangle \right|
=G_{2}
\left|\intop_{-\infty}^{\infty}\,\psi_{f}^{\dagger}\left(\frac{e_{y}\Delta_{y}\sigma_{z}}{\sqrt{\Delta_{y}^{2}+\Delta_{z}^{2}}}\right)\psi_{i}\,dx\right|
\delta_{k_{y,i},k_{y,f}},
\label{eq:keep_sym}
\end{equation}
where $\psi_{i}=\left(\psi_{I,i},\psi_{II,i}\right)$, $\psi_{f}=\left(\psi_{I,f},\psi_{II,f}\right)$, $G_2=G_1/4$, and the indices $i$ and $f$ correspond to the initial and final states respectively. When the transition occurs between states of differing parity, the matrix element of transition, Eq.~(\ref{eq:optical_matrix_element}), becomes
\begin{equation}
\left|\left\langle f\left|\delta H\right|i\right\rangle \right|
=G_{2}
\left|\intop_{-\infty}^{\infty}\,\psi_{f}^{\dagger}\left(e_{x}\sigma_{y}+\frac{s_{\mathrm{K}}e_{y}\Delta_{z}\sigma_{x}}{\sqrt{\Delta_{y}^{2}+\Delta_{z}^{2}}}\right)\psi_{i}\,dx\right|
\delta_{k_{y,i},k_{y,f}}.
\label{eq:exchange_sym}
\end{equation}

We will now analyze the optical selection rules in two distinct regimes. The first regime corresponds to the case where the dispersion lines are linear, here the electron-like branches have positive dispersion, whereas the hole-like branches have negative dispersion. The second regime corresponds to the case where there is a significant reconstruction of the eigenfunctions, i.e., a strong admixture of electron-like and hole-like eigenfunctions (see Eq.~(\ref{eq:elecron_hole_wave1}) and Eq.~(\ref{eq:elecron_hole_wave2}). 

In the linear dispersion regime, one can clearly see from Fig.~(\ref{fig:wave_1}) that the electron-like states are highly localised in the well regions, whereas the hole-like states are highly localised in the barrier region. The overlap between electron-like and hole-like states becomes increasingly small with increasing $\Delta_y$, which leads to all transitions between branches of positive and negative dispersion being heavily suppressed. Similar to non-relativistic quantum wells, in our double well system, within the linear regime, transitions between electron-like (hole-like branches) branches are allowed for light-polarized normal to the direction of the waveguide, while for light-polarized along the direction of the waveguide these transitions rapidly become vanishingly small with increasing $\Delta_y$.

For a massless Dirac system, in the energy range far away from an avoided crossing, but in the region where the splitting between doublet states is maximal, transitions between the two guided modes comprising the doublet are only allowed for light linearly polarized normal to the direction of the waveguide. In stark contrast, in the vicinity of the avoided crossings, transitions strongly occur between the avoided crossing states for light linearly-polarized along the direction of the waveguide. It should also be noted that transitions are also allowed between an avoided crossing level and the state belonging to an ``opposite" parity branch (opposite parity in the large effective mass limit) which lies in between, in this instance transitions are predominately polarized normally to the waveguide.

The energy-level splitting associated with quantum tunneling  between wells and the avoided crossings associated with electron-hole repulsion can be utilized for terahertz (THz) applications. For the appropriate choice of parameters, the potential given by Eq.~(\ref{eq:pot_tobe_solved}) can be used to create doublet states within the vicinity of graphene's charge neutrality point with an energy-level splitting in the THz range. Similarly, the energy gap associated with avoided crossings can be engineered to fall within the THz regime. These avoided crossings would absorb in a narrow frequency range due to the presence of the van Hove singularity at the pseudo-gap edge. A detailed analysis of the optical selection rules shall be a topic of future study, and is beyond the scope of this paper. Our simple analysis serves to demonstrate the potentiality of double-well smooth electron waveguides as the basis of polarisation sensitive THz detectors.



\section{Conclusion}\label{Conclusions}
We have found a class of 1D double-well potentials, for which the 2D Dirac equation can be solved in terms of confluent Heun functions, and calculated the corresponding energy spectra. The energy-level splitting associated with electron tunneling between wells as well the electron-hole avoided crossing gap, can be controlled via the parameters of the electrostatic potential. Dipole optical transitions between the two states comprising a doublet, as well as transitions across the avoided crossing gap are both allowed, but follow drastically different polarization selection rules. For the doublet transitions, light is strongly absorbed for linear polarizations oriented normal to the waveguide. For the avoided crossing transitions, absorption of radiation polarized along the direction of the waveguide dominates. The presence of the van Hove singularity at the bottom of the pseudo-gaps opened in the spectrum leads to a narrow absorption peak, which can be tuned via the applied top-gate voltages to occur in the THz range.

\section*{Acknowledgments} 
This work was supported by the EU H2020 RISE project TERASSE (H2020-823878). RRH acknowledges financial support from URCO (71 F U 3TAY18-3TAY19). The work of MEP was supported by the Ministry of Science and Higher Education of Russian Federation, Goszadanie no. 2019-1246. 

\appendix
\section{Exponentially-decaying double-well}
\label{tanh}

\begin{figure}
\includegraphics[width=0.25\linewidth]{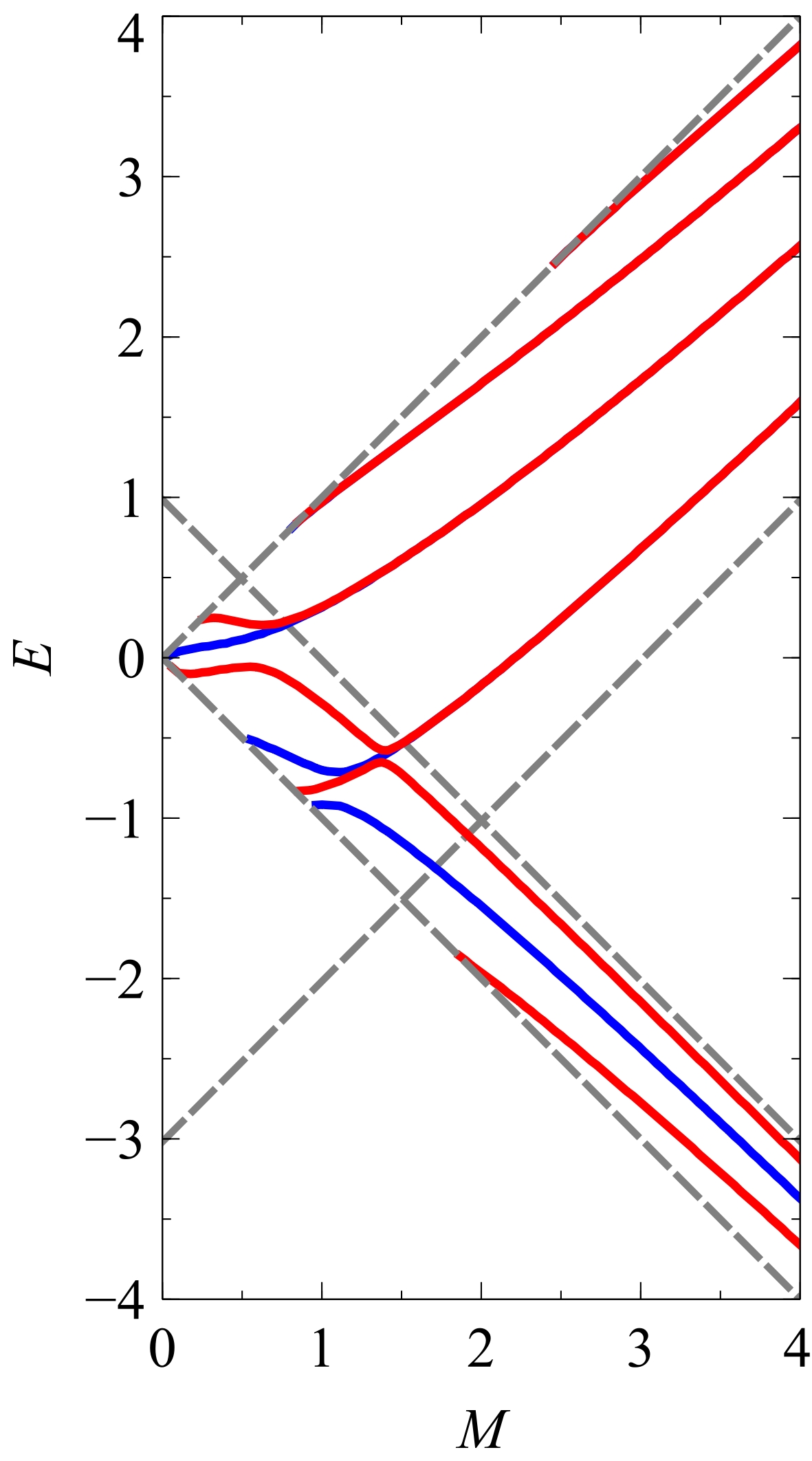}
  \caption{
The energy spectrum of bound states contained within a Rosen-Morse-double-well potential, defined by the potential parameters (a) $B_1=14$ and $B_2=-1$, as a function of effective mass, $M$. Here, only the four lowest doublet states are displayed, while all barrier states are present within the energy range shown. The alternating red and blue lines represent the even (odd) and odd (even) modes of $\psi_I$ $(\psi_{II})$ respectively. The boundary at which the bound states merge with the continuum is denoted by the grey-dashed lines.
}
\label{fig:tanh}
\end{figure}

Although the double-well constructed from the potential Eq.~(\ref{eq:pot_tobe_solved}) is smooth and continuous across the whole of configuration space, the absolute value of the argument of the Heun function appearing in Eq.~(\ref{eq:link}) exceeds unity. Therefore, analytic continuation is needed to analyze its far-field behavior. In contrast, the corresponding Heun function for the Rosen-Morse potential,
\begin{equation}
V=-\frac{B_{1}}{4}\left[1-\tanh^{2}\left(\tilde{x}\right)\right]-\frac{B_{2}}{2}\left[1-\tanh\left(\tilde{x}\right)\right],
\label{eq:RM_Pot}
\end{equation}
is maximally one, and therefore its far-field behaviour can be determined by the expansion about the second pole \cite{hartmann2017two}. Using the coordinate transformation, $\tilde{x}\rightarrow\left|x'\right|-d$, allows the Rosen-Morse potential to model a double-well centred about $\left|\tilde{x}'\right|=0$ . Although this potential is continuous, it is not smooth.  However, in the limit that $\tanh\left(d\right)\approx1$ the discontinuity in its derivative becomes vanishingly small, making it quasi-smooth. This potential has the additional advantage that it can also be used to model shallow double-wells. The potential parameters $B_1$ and $B_2$ appearing in Eq.~(\ref{eq:RM_Pot}) are related to the parameters in Eq.~(\ref{eq:poly_form}) via the relations:
\begin{equation}
p=-is_{p}s_{c}\frac{B_{1}}{4},
\:
\beta=-s_{c}s_{p}\left(s_{j}-i\frac{B_{2}}{2}\right),
\:
\lambda=-\left(is_{j}+\frac{B_{2}}{2}\right)\frac{B_{2}}{2}+2\left(E+\frac{B_{2}}{2}\right)\frac{B_{1}}{4},
\nonumber
\end{equation}
\begin{equation}
s=\frac{B_{2}}{2m}\left(\frac{B_{2}}{2}+E\right),
m=\frac{s_{M}}{2}\left\{ \sqrt{M^{2}-E^{2}}+s_{m}\sqrt{M^{2}-\left(B_{2}+E\right)^{2}}\right\},
\nonumber
\end{equation}
where $s_M,s_m=\pm1$ and the spinor components are given by the expressions
\begin{equation}
\psi_{j}=\sum_{s_{M}}\widetilde{D}_{s_{M}}\left(1-\xi\right)^{\frac{\gamma-1}{2}}\left(1+\xi\right)^{\frac{\delta-1}{2}}e^{-p\xi}u\left(\frac{1-\xi}{2}\right),   
\end{equation}
where $\xi=\tanh\left(\tilde{x}\right)$. As $x\rightarrow \infty$, $\xi\rightarrow1$ therefore $s_M$ must be equal to $1$ for the functions to decay at infinity. The bound states are obtained via the zeros of the functions $\psi_{I}(\tilde{x}'=0)$ and $\psi_{II}(\tilde{x}'=0)$. In Fig.~(\ref{fig:tanh}) we plot the energy eigenvalue spectrum for the potential, Eq.~(\ref{eq:RM_Pot}), defined by the potential parameters $B_1=14$ and $B_2=-1$, displaying the four lowest doublet states and complete set of barrier states within the energy range displayed.

\section{Lorentzian Potential}
\label{sec:Lorentzian}
\begin{figure}[ht]
\includegraphics[width=0.25\linewidth]{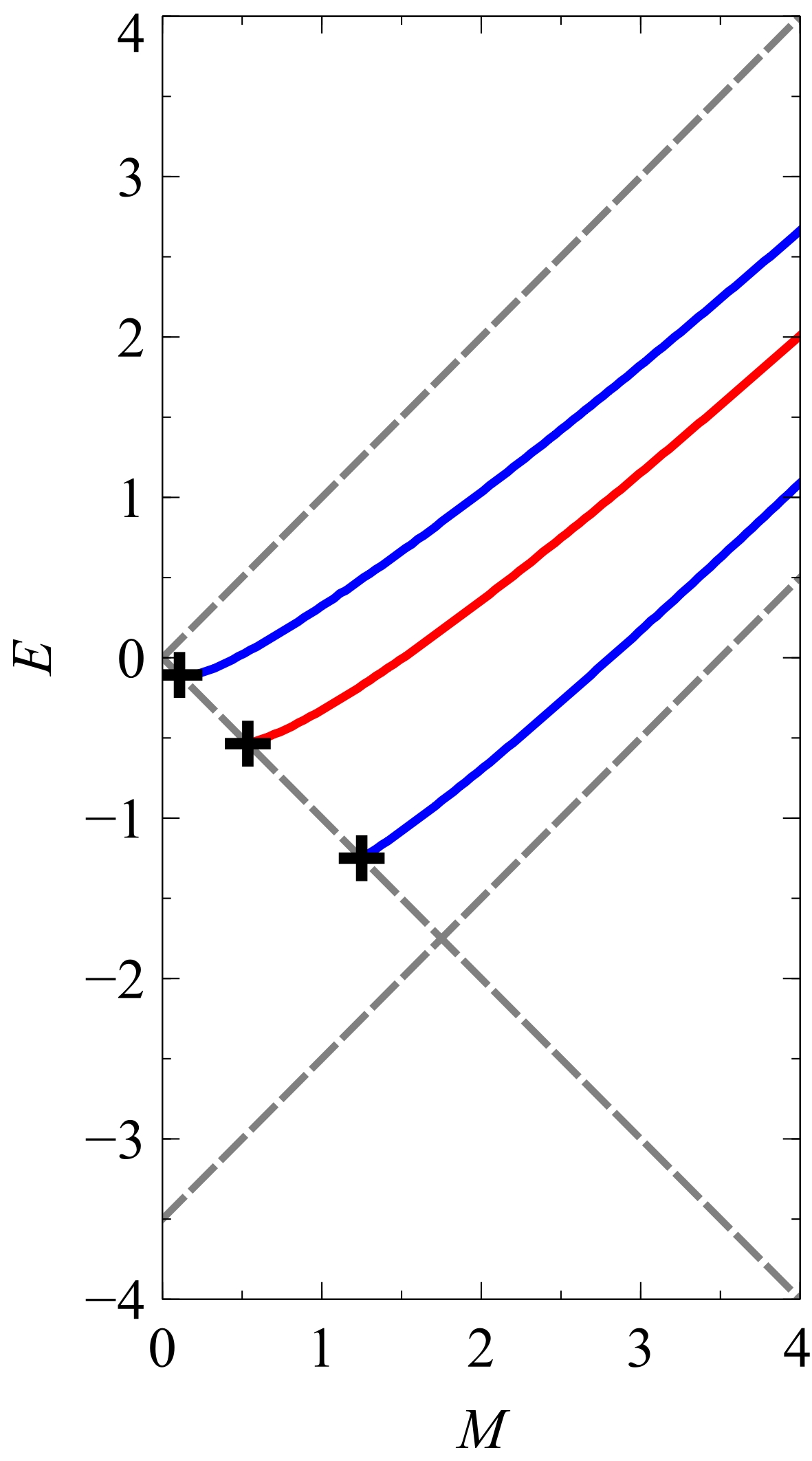}
  \caption{
The energy spectrum for the three lowest-energy guided modes contained with a Lorentzian potential, of strength $A_{0}=-3.5$, as a function of $M$. The alternating red and blue lines represent the even (odd) and odd (even) modes of $\Psi_{I}$ ($\Psi_{II}$) respectively. The black crosses denote the supercritical states. The boundary at which the bound states merge with the continuum is denoted by the grey-dashed lines.  
}
\label{fig:Lorentz}
\end{figure}

When $p=0$, i.e. $\left| E\right|= M$, and $A_{1}=0$, the Heun confluent function appearing in Eq.~(\ref{eq:link}), reduces to a Gauss hypergeometric function, therefore
\begin{equation}
u\left(\widetilde{\xi}\right)=\left(1-\widetilde{\xi}\right)^{-\omega} {}_{2}F_{1}\left(-s_{a}\left(s_{c}s_{j}+A_{0}\right)+\omega,\,\omega;\,\gamma,\frac{\widetilde{\xi}}{\widetilde{\xi}-1}\right),
\label{eq:hypergeom}
\end{equation}
where $\omega=s_{a}A_{0}+\left(1+s_{\omega}\sqrt{1+8EA_{0}}\right)/2$, and ${}_{2}F_{1}$ is the hypergeometric function. The particular case of $A_{1}=0$ corresponds to the Lorentzian potential, which notably admits exact eigenvalues for supercritical states. The emergence of a supercritical state (i.e. bound states whose energy, $E=-M$), is characterized by the appearance of a new node at infinity~\cite{hartmann2017two}. As $\tilde{x}'\rightarrow\infty$, $\psi_{j}$ takes the form of
\begin{equation}
\lim_{x\rightarrow\infty}\left(\psi_{j}\right)\propto R^{\frac{m+s+1}{2}}\left[\sum_{s_l}\left(1-R\right)^{-\frac{1+\Omega_{l}}{2}}\frac{\Gamma\left(1+s_{a}A_{0}-s_{a}s_{j}s_{c}\right)\Gamma\left(\Omega_{l}\right)}
{
\Gamma\left(s_{a}A_{0}+\frac{1+\Omega_{l}}{2}\right)
\Gamma\left(\frac{1+\Omega_{l}}{2}-s_{a}s_{c}s_{j}\right)
}
\right],
\label{eq:lorent_farfield}
\end{equation}
where $\Omega_{l}=s_{l}s_{\omega}\sqrt{1+8EA_{0}}$, $s_l=\pm 1$ and $R=\widetilde{\xi}/(\widetilde{\xi}-1)$. It can be seen from Eq.~(\ref{eq:lorent_farfield}) that the bound state condition is contingent upon the value of $\Omega_l$. For the case of $\Im\left(\Omega_l\right)=0$ and $\left|\Omega_l\right|>1$, bound states arise when 
\begin{equation}
E=\frac{A_{0}^{2}+s_{a}\left(2n+1\right)A_{0}+n\left(n+1\right)}{2A_{0}},
\nonumber
\end{equation}
where $s_{a}A_{0}<0$ and $n=0,1,2,\ldots$ and $n$ is restricted such that 
$n<\left|s_{a}A_{0}\right|-1$.
In Fig.~(\ref{fig:Lorentz}) we plot the energy eigenvalue spectrum for the case of the Lorentzian potential with potential strength $A_{0}=-3.5$. The exact supercritical eigenvalues are indicated in black crosses.


\bibliography{ref}

\end{document}